\documentclass{article}

\usepackage[english]{babel}

\usepackage[letterpaper,top=2cm,bottom=2cm,left=3cm,right=3cm,marginparwidth=1.75cm]{geometry}

\usepackage{amsmath}
\usepackage{graphicx}

\usepackage{caption}
\usepackage{subcaption}
\usepackage{natbib}
\usepackage{float}
\usepackage[colorlinks=true, allcolors=blue]{hyperref}
\usepackage{xcolor}
\usepackage[final]{changes}

\title{Words are not Wind - How Public
Joint Commitment and Reputation Solve the Prisoner's Dilemma
}
\author{
Marcus Krellner$^{1,2,\star,a}$ and The Anh Han$^{1,b}$}

\begin{document}
\maketitle
{\footnotesize
		\noindent
		$^{1}$  School of Computing, Engineering and Digital Technologies, Teesside University, UK\\ 
            $^2$ School of Mathematics, University of St. Andrews, UK\\
		$^\star$ Corresponding authors: krellner.marcus@gmx.de\\
            $^a$ https://orcid.org/0000-0001-8269-965X\\
            $^b$ https://orcid.org/0000-0002-3095-7714
	}

\begin{abstract}
To achieve common goals, we often use joint commitments. Our commitment helps us to coordinate with our partners and assures them that their cooperative efforts will benefit themselves. However, if one of us can exploit the other's cooperation (as in the Prisoner's Dilemma), our commitment appears less useful. It cannot remove the temptation for our partners  to exploit us. Using methods from evolutionary game theory, we study the function of joint commitments in the Prisoner's Dilemma. We propose a reputation system akin to indirect reciprocity, wherein agents observe interactions even when not directly involved. They judge cooperation as good and defection as bad, but, crucially, only if the parties involved had committed to cooperate. This results in stable cooperation even though judgments are made privately, which had been a weakness in previous models of indirect reciprocity. Our work shows that joint commitments have utility beyond coordination problems, which could explain their prevalence. The proposed link between joint commitments and reputation could also explain why some joint commitments are pointedly public, like wedding vows. A reputation-based mechanism might have been particularly relevant in our distant past, in which no institutions existed to enforce commitments.

\deleted[]{To achieve common goals, we often use joint commitments. Our commitment helps us to coordinate with our partners and assures them that their cooperative efforts will benefit themselves. However, if one of us can exploit the other's cooperation (as in the Prisoner's Dilemma), our commitment appears less useful. It cannot remove the temptation for our partners  to exploit us. 
Using methods from evolutionary game theory, we study the function of joint commitments in the Prisoner's Dilemma. We propose a reputation system akin to indirect reciprocity, wherein agents observe interactions even when not directly involved. They judge cooperation as good and defection as bad, but, crucially, only if the parties involved had committed to cooperate. 
This results in stable cooperation even though judgments are made privately, which had been a crucial weakness in previous indirect reciprocity models. Our work shows that joint commitments have utility beyond coordination problems, which can explain their prevalence. The proposed link between joint commitments and reputation can also explain why some joint commitments are pointedly public, like wedding vows. A reputation-based mechanism might have been particularly relevant in our distant past, in which no institutions existed to enforce commitments.}
\end{abstract}


\section{Introduction}

The Prisoner's Dilemma (PD) \citep{Sigmund,Doebeli2005} has been attracting the interest of researchers for decades \citep{Rapoport1965,Xu2025}, because it intuitively explains and precisely models the problem of cooperation. In this game, two individuals can work together for mutual benefit. But if one defects, while the other cooperates, the defector has the highest benefit, while the cooperator receives the worst possible outcome. This leads both parties to defect, an outcome that is worse for either party than if they had chosen to cooperate. The classic scenario of the game involves the namesake prisoners, who need to decide to keep their silence (cooperate) or betray incriminating information for a lenient sentence (defect). But the Prisoner's Dilemma represents countless other situations with the same underlying payoff structure \citep{Poundstone2011}.  

The problem has been studied extensively with classic and evolutionary game theory. The latter investigates under which conditions cooperation can evolve, i.e. when cooperators can earn higher payoffs than defectors and supplant them in a population. Such models are essential to understand why humans cooperate at all and under what conditions they fail to do so. So far, a few general principles have been discovered that make it possible: kin selection \citep{Hamilton1964} and group selection \citep{Traulsen2006}, direct and indirect reciprocity \citep{Nowak2006,Nowak2005}, spatial networks \citep{Szabo2007,Perc2013}, and different forms of incentives \citep{Fehr2000,Sigmund2001}. Of them, indirect reciprocity stands out because it allows cooperation in unrelated, 
unstructured populations with little chance of repeated interactions \citep{Rand2013,Nowak2005,Okada2020:review}. This condition becomes increasingly important in a world in which billions of people are getting more and more connected. It is no surprise that real-world applications of indirect reciprocity, namely reputation-based systems, are prevalent in e-commerce systems \citep{Shouxu2007,Standifird2001}.


Another solution with similar potential has been proposed in the form of arrangements \citep{Han2012:com,Han2013:agreements}, also called joint commitments. Before playing the PD, the two parties can promise to cooperate and agree on a compensation that would punish defection and/or reward the cooperation. Although being a valid system, this approach relies on the enforceability of such compensation. When the time comes, the defector will be reluctant to pay. Implementing the compensations might require some form of powerful governing entity that intervenes itself or provides incentives \citep{Han2022}. Our modern states with their judiciary systems are capable of enforcing joint commitments, especially written contracts. But we should also be interested in how joint commitments functioned before these modern accomplishments, or in interactions that are too insignificant for a state to intervene in (e.g. a promise to meet for a coffee). 

To date, mere joint commitment, i.e. the mutual promise of cooperation without compensation, has never been demonstrated as a solution to the PD with the methods of evolutionary game theory. Showing that would be important for multiple reasons. First, such joint commitments are often considered as a solution to the PD. The dilemma of the prisoners seems easier if we allow them to assure each other that they will cooperate. In fact, the thought experiment is phrased so that the prisoners are prevented from talking with each other \citep{Rapoport1989}. 
Second, we cannot explain why mere commitment would actually make a difference. If prisoner A claims to cooperate and prisoner B believes them, it is still in the best interest of B to defect. In addition, because A would benefit from B's cooperation, A has an incentive to lie to B. So B should not believe A in the first place. Third, joint commitments, without guaranteed compensations, are so ubiquitous that they “make our social world” \citep{Gilbert2014:JCbook}. Humans learn their use and importance at a very young age \citep{Kachel2019,Grafenhain2013,Kachel2018,Grafenhain2009,Kachel2019a}. Lastly, how this prevalence was explained so far, might be wrong. Researchers have argued that joint commitments are mostly used to solve coordination problems \citep{Tomasello2012,Tomasello2019}. Such problems are simpler than the PD \citep{Forber2015}. If we could show that joint commitment can also solve the PD, we could fundamentally change the understanding of the ultimate function \citep{Tinbergen1963} of joint commitment, with huge impacts on philosophy, psychology, and anthropology. 


We will first elaborate on the concept of joint commitments. 
A commitment is a promise to do something\footnote{\added{The phrase “joint commitment” is not always used in the same way. We will use a notion related to 'making a joint commitment', which always includes the utterance of some promises. This is in line with many scholars }\citep{Clark2020,Shpall2014}\added{. However, as our model will demonstrate, commitments are more complex than promises alone. Promises create obligations. This is more related to the notion of ‘being committed’. We include both, the promise and the obligation, in our model. Other scholars used the phrase ‘being jointly committed’ to describe any emergence of obligations, even without a promise }\citep{Michael2016,Gilbert2018:remarks}\added{. Although related, our model cannot be directly applied to these ideas.}}\added{. \textit{Joint} commitment entails that at least two people make a promise. But it is a peculiar kind of promise, because it is only valid if both parties are willing to commit. A good example is marriage vows. Both people are asked 'Do you want?' and only if both answer 'Yes!' a joint commitment is proclaimed, and they are now married. If in this moment only one person says yes and the other says no, the first promise is without consequences. The person who said yes would not be part of a joint commitment (that is, the marriage), even if they have already promised.}

\begin{figure*}[!htb]
\begin{center}
\includegraphics[width=0.95\linewidth]{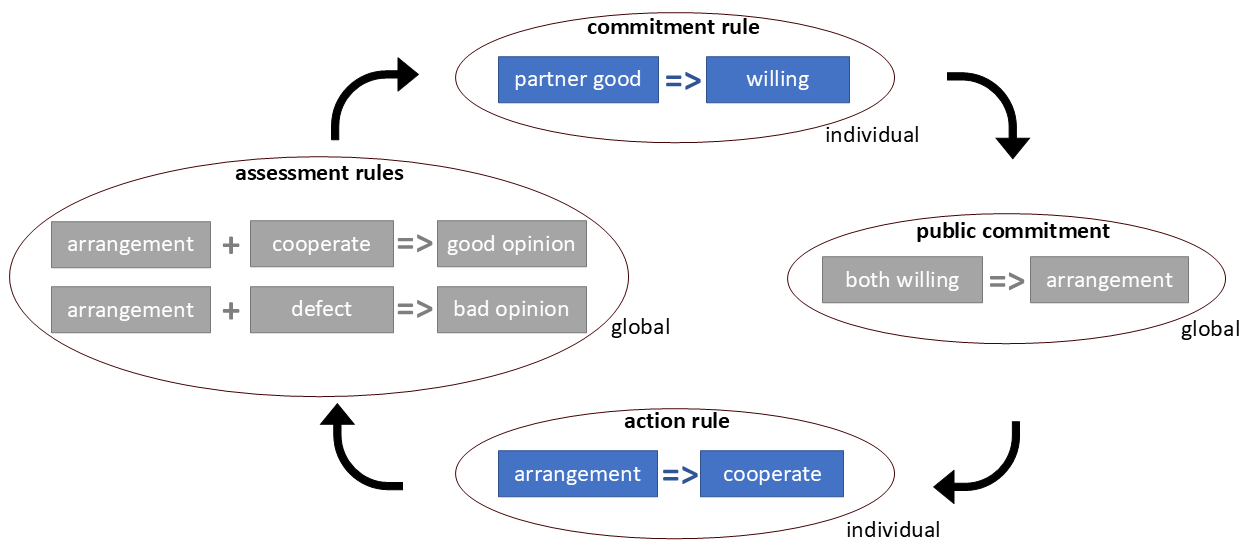}
\caption[Principles of Joint Commitment]{Principles of Joint Commitments. \textmd{This schema shows all the steps of a strategy that uses joint commitment and reputations to enable cooperation. When  playing the Prisoner's Dilemma (PD) with another individual, you first apply the commitment rule (top). If you think your partner is good (i.e. trustworthy), you are willing to enter a joint commitment. If you are both willing, a joint commitment (or “arrangement”) is made. In such an arrangement, you will cooperate. Everybody assesses actions that are made in an arrangement. The opinions formed will be used when they play the PD with this individual in the future. The symbol '$=>$' indicates 'if and only if', so that rules could be simplified (i.e. these rules were not listed: partner bad $=>$ not willing, no arrangement $=>$ defect, no arrangement + any action $=>$ no assessment.)}

\textmd{For later reference, we indicated some steps as individual (blue) or global (grey). This distinction will be referred to in the Method section when we introduce the evolutionary setting.
}}
\label{fig:principles}
\end{center}
\end{figure*}

We use this form of joint commitment and combine them with reputations in a new model, illustrated in Figure \ref{fig:principles}. Two individuals meet randomly and play the PD. They consider the reputation of their partner. They are only willing to enter a joint commitment if they trust them.  Only if both parties signal their trust, they publicly declare their joint commitment to cooperate. In this scenario, they keep their commitment and cooperate in the PD. If either party is unwilling and no joint commitment is made, they choose to defect. Others observe these actions. By cooperating in a joint commitment, an individual earns a good reputation, which fosters trust in future partners. However, if a partner fails to cooperate within a joint commitment, they lose trust. If no joint commitment occurs, neither party is judged. In a large population with a reputation system, individuals frequently encounter partners for the first time. However, they can discern trustworthiness based on others' interactions.

\added{
Our model is closely related to some previous approaches, yet differs in several important aspects. The reputation system is in many ways similar to the ones that have been studied extensively for so-called indirect reciprocity }\citep{Nowak1998:IRinNature,Ohtsuki2006}\added{, for reviews see }\cite{Nowak2005,Okada2020:review}\added{. They studied different assessment rules that determine what makes someone worthy of cooperation. Our assessment rules instead determine what makes a person worthy of trust. Trust in turn determines whether individuals enter joint commitments, and commitment determines if they cooperate. We will show how this intermediate step has the potential to resolve the problem of disagreement} \citep{Panchanathan2003:importance-standing,Krellner2022}, which has been a long-standing issue in the research on indirect reciprocity \citep{Uchida,Hilbe2018:PriA,Radzvilavicius2019,Okada2020:TwoWays,Perret123,Krellner2023:agr}. \added{We will highlight this point in the method section.
}

\added{In this study, we tested our hypothesis that a strategy integrating  joint commitments and reputations can promote the evolution of cooperation within a population of self-interested individuals. We use agent-based simulations for the evolutionary process and perform analytical predictions and simulations for reputation dynamics. Our results show that the proposed strategy is highly prevalent in evolutionary dynamics, leading to high levels of cooperation. We }
\deleted[id=.]{In the remainder of this article, we will elaborate the model, including formal definitions of possible strategies. Note that we will use 'joint commitment' and 'arrangement' interchangeably throughout the Methods and Results sections, as they mean the same in this content. We will then show how long-term behaviour can be analytically predicted and validate these predictions with simulations of the game. We will use our predictions to calculate expected payoffs for a selection-mutation process in a Monte Carlo simulation to study the evolution of cooperation. Finally, we will}conclude with a detailed discussion of our results and their implications.

\section{Models and Methods}

\subsection{Model}
\label{ch:2.1}

\begin{figure}[H]
\centering
  \includegraphics[width=0.65\linewidth]{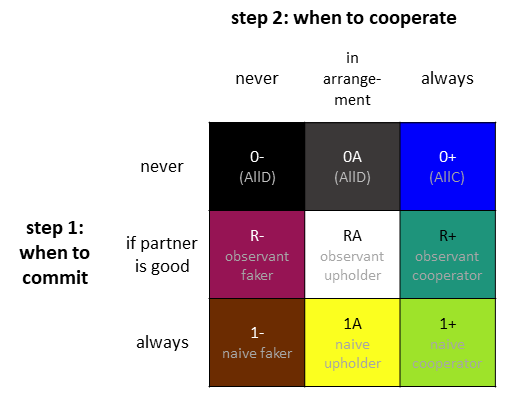}
  \caption[Strategies]{Strategies\textmd{, i.e. combinations of commitment rules, $\alpha$, and cooperation rules, $\beta$. Strategies are named by abbreviation, such as RA for '\textbf{R}eputation observant \textbf{A}rrangement upholder'. They broadly fall into three categories: the mentioned RA which applies all principles, 'nice' strategies (yellow, green and blue) which are either more generous (cooperate outside arrangement), more naive (enter arrangements with everybody) or both, and lastly 'mean' strategies (dark and red) that either never commit nor cooperate (AllD) or commit but defect anyway (fakers).}}
  \label{fig:strats}

\end{figure}

\added{This section consists of three parts. First, we provide a detailed account of our model, including the structure of the population, the sequence of the game, and all possible strategies. Second, like all reputation-based models, we derive the average reputations and cooperative behaviour associated with each strategy. Third, we use these averages to derive the long-term evolutionary process. Note that, throughout the Methods and Results sections, we will use 'joint commitment' and 'arrangement' interchangeably, as they hold the same meaning in this context.}

We consider a well-mixed population of $N=100$ individuals who interact with each other via a version of the Prisoner's Dilemma game. The game payoffs are defined by two parameters: benefit ($b$) and cost or cooperation (which is set to 1); see Figure \ref{fig:principles}. In each round, two randomly selected players ($i,j$) play the Prisoner's Dilemma. Each round is played in three steps: commitment, cooperation, and assessment. \added{A strategy is defined by behavioural rules for each  step. We will compare the proposed strategy, RA, with previously introduced reputation-based strategies for a clearer understanding. We will focus on the Image Scoring \citep{Nowak1998:IRlong,Nowak1998:IRinNature}}, more precisely, the simplified version that uses only the reputations 'good' or 'bad'; and the Staying strategy \citep{Okada2017,Sasaki2017,Okada2018}, since they are widely acknowledged and informative, in this context.\footnote{We will describe the relevant features of these strategies later, but a minimal summary is as follows. Image Scoring considers all defection as bad and all cooperation as good. Whereas, Staying considers only defection against good players as bad and only cooperation with good players as good, while it ignores all interactions with bad players. Both cooperate if they consider the partner to be good, and defect if they consider them bad.} 

In the first step, both players individually decide whether they offer to commit, using their commitment rule ($\alpha$). \added{There are three version of this rule, see rows in Figure \ref{fig:strats}: always commit (strategies that start with "1"), never commit ("0"), or only offer to commit only when you consider your partner as good ("R")}. If both offer to commit, an arrangement is made\added{ and publicly acknowledged} ($a=1$); otherwise it is not ($a=0$), even if one of them offered commitment. \added{If an arrangement is made, both players need to pay the cost $c_a$. This represents any necessary or voluntary costs of the arrangement (for example, marrying in an expensive location). }Players then make their choice in the Prisoner's Dilemma ($x_i, x_j$), to cooperate (\added{$x=1$}) or to defect (\added{$x=0$}) using their cooperation rule, denoted by $\beta$. It can be unconditional defection (\added{strategies that end with "-"}), unconditional cooperation ("+"), or conditional cooperation ("A"), which occurs only in the presence of an arrangement (see the three columns of Figure \ref{fig:strats}). \added{If they cooperate, they pay the cost of cooperation ($c=1$) and their partner gains the benefit $b$.} 

We will illustrate the resulting payoffs with three examples. If RA meets another RA, let us assume that they trust each other. Hence, an arrangement is made and $c_a$ occurs. They will then both cooperate, so they lose $c$ but gain $b$. Their payoff is $P=-c_a-c+b$. If the RA meets a 1- instead (who they also trust), the arrangement is still made, but their partner defects. They therefore endure both costs without reward $P=-c_a-c$. If they meet a 0- instead, an arrangement cannot be made, since this partner is not willing to enter one. The resulting payoff is $P=0$. The definition of each pairwise payoff, which considers the probability of trust between each pair, will be given in Section \ref{ch:2.3}.

\added{When we compare a RA-player using our proposed strategy, with players  using previously proposed strategies, we observe both similarities and differences. Of course, neither of the previous strategies used commitment, so they would never commit. But who would they cooperate with? Both Image Scoring and Staying will cooperate if they consider their partner as good and defect when they consider them as bad. In the same way, an RA-player will always defect when they consider their partner as bad, since they will never join commitments with such partners. However, RA will also defect if no joint commitment came about for other reasons, for example if their partner just never commits or considers the RA-player as bad.}

In the third step, after the two interacting players have made their choices, everyone (including themselves) will assess them and potentially update their opinion. Opinions can only be good or bad, and everyone has an opinion about everyone. The opinions are stored in a so-called 
image matrix \citep{Uchida}. \deleted[id=.]{ We call the average opinion about a player their reputation $r$.} \added{Assessments can have three forms: someone} likes an action, dislikes it, or feels neutral about it \citep{Krellner2022}. If the opinion is currently bad and the action is liked, the opinion changes to good. Similarly, if the opinion is good and the action is disliked, it changes to bad. Otherwise, the opinion remains the same. \added{For simplicity, we consider all strategies to be initially trusting and therefore initiate the image matrix with good opinions. Since we are interested in long-term behaviour, the initial state is less important. But see also SI for further discussion.} 

\added{In this study, we use the following way of assessment. If there is no arrangement, all actions are neutral. If there is an arrangement, cooperation is liked, and defection is disliked. Crucially, all players may make mistakes when they assess. In particular, with probability $\epsilon$, they make perception errors in which they mistake a cooperation for a defection, or vice versa. Each observer makes these mistakes independently of other observers and independently of each player they observe.}

\deleted[id=.]{To assess, players use a norm $\gamma$ citep{Ohtsuki2004a,Okada2020a}.A norm consists of four rules ($\gamma_{11},\gamma_{10},\gamma_{01},\gamma_{00}$) to take into account the presence of arrangement ($\gamma_{1j},\gamma_{0j}$) and if the assessed player cooperated ($\gamma_{i1},\gamma_{i0}$). The context of the arrangement replaces the context of the reputation of the partner (or recipient) from previous reputation models.}

\added{Again, our proposed strategy is slightly different from previous ones. Someone using the Image Scoring norm would assess all cooperation as good and all defection as bad, regardless the circumstances. The Staying norm applies a more complex judgement. If the observer considers the partner of someone as bad, the observer will ignore their action.}
\added{But if they consider the partner as good, they apply the same rules as Image Scoring . This is very similar to our proposal, which ignores actions outside arrangements and applies Image Scoring inside arrangements. Both ways of assessing consider a single additional circumstance. However, the important difference is the control over this circumstance. A player cannot decide what opinion the observers currently have about their partner. But the player can decide whether an arrangement is made. Therefore, in our model, an observer’s assessment of a player cannot be negatively influenced by the observer’s current opinions of the players. Assessments are therefore not influenced by disagreements between active players and observers.}

\deleted[id=.]{As an example, consider the RA strategy with the assessment rule $\gamma=\{1,-1,0,0\}$. This strategy commits only if its partner is good and cooperates only if an arrangement was made; and if an arrangement was made, it judges cooperation as good and defection as bad, and does not care about what happened otherwise (i.e. if no arrangement was made). This RA strategy can be considered as '(reputation) observant (arrangement) upholder', see Figure \ref{fig:strats}. It fits our earlier deliberations on what a commitment strategy should look like. We will focus our investigation on nine different combinations of commitment and cooperation rules (i.e., nine 'strategies') for the norm $\gamma=\{1,-1,0,0\}$. This norm closely resembles the famous "Staying" norm \citep{Okada2017,Sasaki2017}, which is also known as the number seven (L7) of the leading-eight \citep{Ohtsuki2006}. The reason for focussing on this norm will become apparent in the next section. }

\subsection{Analytical Prediction of Reputations}

\added{We refer to  the average opinion about a player as their reputation, denoted by $r$. For example, a player whom 90 other players consider to be good and 10 to be bad has a reputation $r=0.9$. Keep in mind that in our model, all nine strategies assess in the same way. }For \deleted[id=.]{all strategies of the norm ($\gamma=\{1,-1,0,0\}$)} this way of assessment, we can predict the long-term average reputation by just considering the rate of perception errors. This is a known property of reputations for some norms in limited circumstances \citep{Fujimoto2022}. It was first demonstrated for the \deleted[id=.]{(simple)} \added{mentioned} Image Scoring norm \citep{Nowak1998:IRlong}. In Image Scoring, under full observation, the resulting reputation of a player is independent of their former reputation. It only depends on whether they cooperated or not.
and what percentage of observers made a perception error $\epsilon$. After cooperation, all observers who perceived correctly will have a good opinion, all that made an error will have a bad one. In an infinitely large population, the reputation of a player can therefore only be $r=1-\epsilon$ (after cooperation) or $r=\epsilon$ (after defection).

This principle also applies to the current norm. The assessment rules for the two cases with arrangement are equal to the Image Scoring norm: good if cooperated, bad if defected. And in the two other cases with no arrangement, reputation never changes. \deleted[id=.]{The strategies are slightly different, though. In Image Scoring, the choice to cooperate depends on the opinion about the partner. They only cooperate with good partners. Hence, a well-meaning Image Scoring player will end up with a low reputation ($r=\epsilon$) if they disliked their partner. A well-meaning RA does not face this problem. They can only be judged if they enter an arrangement, and if they did, they will always cooperate.} 
\added{Given the similarities, we can, in principle, predict the long-term average of reputations of all our strategies in infinite populations. However, since our strategies are more complex than Image Scoring, we need to make the following two assumptions.} \deleted[id=.]{But in order to make precise predictions about the reputations in our norm, we need to make two  assumptions. First}

\added{\textbf{Assumption 1:}} the information that an arrangement was made is public \added{(known to all)} and not noisy (i.e. there is no additional type of error which changes the perceived status of arrangement). \deleted[id=.]{Second}

\added{\textbf{Assumption 2:}} at least sometimes, all players are able to enter arrangements, so they can be judged. \deleted[id=.]{As we will see, we will need further clarifications to fulfil the latter statement.}

\begin{figure}[H]
\centering
  \includegraphics[width=0.95\linewidth]{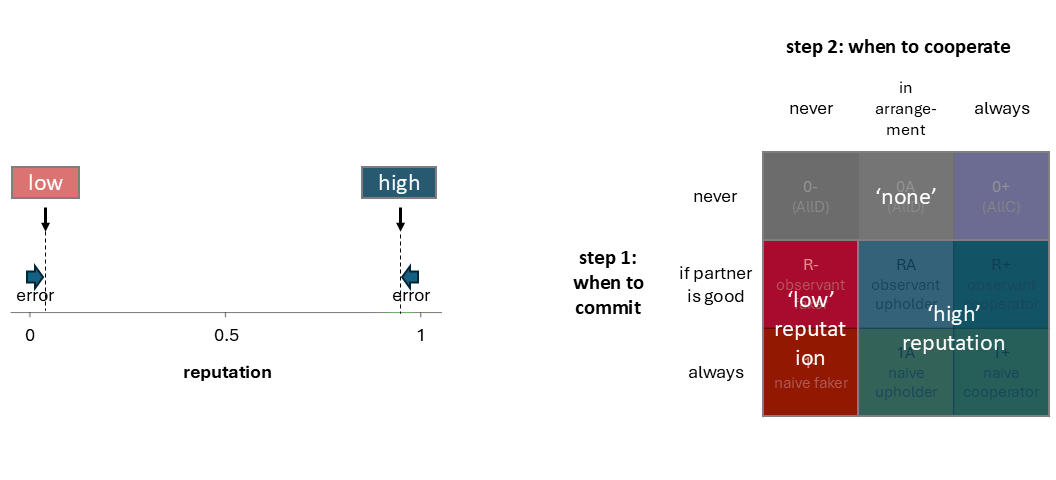}
  \caption[Predictions]{Predictions of Reputations\textmd{. On the right is an overlay of the three different predictions under Assumptions 1 and 2 for the nine strategies. On the left is a graph of reputation values. We predict that faker strategies should always have  bad reputations ($r=0$), but that perception error increases their average reputation to $r=\epsilon$, as is indicated by the broken line. The case is similar for 'high' reputations.}}
  \label{fig:preds}

\end{figure}

We can group the nine strategies into three groups of similar predictions (Figure \ref{fig:preds}). R- and 1-, the so-called faker strategies, will have average reputations of $r=\epsilon$ (prediction 'low'). They will enter arrangements, but then they will always defect. Hence, only observers who make perception errors believe them to be good (recall that the perception error changes defection to cooperation and vice versa in the eye of a particular observer). This means that after their first ever interaction in an arrangement, their reputation in infinite populations is always exactly $r=\epsilon$, not just on average. \deleted[id=.]{Lastly, we predict that all other strategies} RA, 1A, R+ and 1+ always cooperate or at least always cooperate in an arrangement. \added{Therefore they will have reputations} $r=1-\epsilon$ (prediction 'high'). The last group consists of all strategies that never commit, 0-, 0A, and 0+. They are a special case because they themselves refuse to commit, thus never entering arrangements. Hence, they will never be judged (prediction 'none'), so their reputations remain as they were in the beginning. However, we could even ignore the reputations for these strategies. Recall that reputation neither influences whether somebody cooperates nor how they are assessed. It only determines if arrangements are made (which then influences cooperation and assessment). But these strategies would never enter arrangements. Their reputation never affects this fact.

With these predictions for the three groups, we could subsequently predict average rates of arrangements, cooperation, and average payoffs if the population size and timescale are infinite. In finite settings, on the other hand, the reputations will not be exact at the predicted value but scattered around it in a normal distribution \citep{Fujimoto2022}. We could ignore this noisy property, since we are just interested in averaged outcomes over long periods of time. However, finite populations pose yet another problem. In this case, it is possible that the reputation of a player drops to zero. For example, if they defected and, by chance, no observer made a perception error. \added{This gives rise to a scenario that violates Assumption 2. Consider the following combination of circumstances. }First, a player's reputation is $r=0$\added{, nobody trusts them. Second,} there are no \added{players that would enter an arrangement even if they don't trust them, i.e. there are no players of strategy 1-, 1A or 1+). In these circumstances, the player with reputation zero never again gets the chance to enter another arrangement.} \deleted[id=.]{1* strategy players in the population, no other player will enter an arrangement with them ever again, which violates Assumption 2. Therefore, depending on the strategies they face, a player that drops to $r=0$ may} \added{They} do not get any chance to redeem themselves. We need to alter our predictions for these cases. \added{For finite populations with no players of strategy 1-, 1A or 1+, we do not consider Assumption 2 to hold, but instead consider three alternatives:}\deleted[id=.]{Let us consider a version of our original assumption as} 

\added{\textbf{Assumption 2.a}}: over a short period of time, the probability of a player to reach $r=0$ is small enough, so that the predictions 'none', 'low' and 'high' still hold, even if there are no 1*  players. \deleted[id=.]{But let us consider another a}

\textbf{Assumption 2.b}: after an intermediate period of time, all faking players (1- and R-) will reach $r=0$ (the prediction 'zero' holds instead). \added{These players' reputations are most likely to drop to zero. It will happen when} they had entered an arrangement in which no observer made a perception error. The chance of that, for example when there are $N_{o}=50$ observers and the perception error is $\epsilon=0.05$, is given by \deleted[id=.]{the chance to reach $r=0$ is} $p=(1-\epsilon)^N\approx0.08$, for each entered arrangement. 

\deleted[id=.]{Let us finally consider a}\textbf{Assumption 2.c:} after \deleted[id=.]{infinite}\added{a long period of} time, all but 0* strategies (which stay 'none') will reach $r=0$. For 1+, 1A, R+ or RA \added{to reach $r=0$, all observers must make a perception error during the same arrangement. If} $N_{o}=50$ and  $\epsilon=0.05$\added{, the probability for that} is $p\approx10^{-66}$, for each entered arrangement. 

\added{It is important to note that we do not claim to make precise predictions about the reputations under these special conditions. Our assertion is that these represent the three extreme cases. Assumption 2.a is the default case. Assumption 2.b makes it easier to avoid fakers (their reputation is 'zero' instead of 'low'). Assumption 2.c makes cooperation between RA impossible (since they do not trust each other). We do not know which of these cases will occur for a given setting or if there are mixed states somewhere in-between the extremes. However, we avoid the need for precise predictions, by considering each of these extreme cases for our analysis (see Figure \ref{fig:res1}). As we will show, the results remain consistent across all extreme cases.}

\subsection{Simulation of an Evolutionary Process}
\label{ch:2.3}

The predictions of reputations allowed us to predict payoffs and use a Monte Carlo simulation to study an evolutionary process with social learning and mutation. We used finite populations $N=100$ and initiated them in the setting we considered the hardest: all players use strategy 0-, so there is initially no commitment, no cooperation. \added{(We also ran a number of simulations with other initial conditions, such as random distribution of strategies. They showed no difference in long-term behaviour)}. We ran a simulation for $10^5$ turns. Each turn, a mutation event may occur in which a random player changes their strategy at random. (For most results below, mutation probability is set to $\mu=0.01$ \citep{Wu2012,Hilbe2018:PriA,Krellner2021}\added{, but we also used $0.002$ and $0.05$}, as shown in Figure \ref{fig:res1}.) If there was no mutation, a social learning event would occur instead, where a random player (learner) adopts the strategy of another random player (model) with the probability determined by their payoff difference and the Fermi-function (see equation \ref{Fimi} below).

\deleted[id=.]{The average payoff of a player is determined by its pairwise payoff with a player of each strategy and the abundance of that strategy. Pairwise payoffs are combined in the payoff matrix $P$.} \added{Payoffs are dependent on the strategies' frequencies  and the pairwise payoffs in matrix $P$. }A single entry $P_{ij}$ determines what a player of strategy $i$ earns on average in an encounter with strategy $j$. \added{This is determined by} their probabilities to be willing to commit, $w_{ij}$ and $w_{ji}$\added{, the probability of an arrangement, $a_{ij}=w_{ij}$$w_{ji}$ (which is the chance that both are willing),} and their probability to cooperate $x_{ij}$, $x_{ji}$ (which may depend on arrangement), multiplied respectively by the cost of arrangement $c_a$, cost of cooperation $c=1$ and benefit of cooperation $b$. \added{Note that these payoffs are determined by the strategies of the players involved, given by their commit rule $\alpha$ and cooperation rule $\beta$, as well as the previously predicted reputations of the strategies $r$, which are dependent on the perception error $\epsilon$. The payoffs therefore incorporate the entire model.}

\begin{equation}
    P_{ij}= -c_a w_{ij} w_{ji} - c x_{ij} + b x_{ji}; 
   \ \text{where } w_{ij}=
    \begin{cases}
     1, & \text{if}\ \alpha_i=1, \\
     r_j, & \text{if}\ \alpha_i=R, \\
     0, & \text{if}\ \alpha_i=0,
      \end{cases} \text{ and }
    \ x_{ij}=
    \begin{cases}
     1, & \text{if}\ \beta_i=1, \\
     a_{ij}, & \text{if}\ \beta_i=A, \\
     0, & \text{if}\ \beta_i=0.
      \end{cases}
      \label{all}   
\end{equation}

There are $q$ strategies present in a population, where each is adopted by $n_i$ individuals ($N=\sum_{i=1}^q n_i$). The average payoff $\widehat \pi _i$ of the strategy $i$ is therefore

\begin{equation}
    \widehat \pi _i = \frac{1}{N-1}\sum_{j=1}^q P_{i,j} n_j^*,
   \quad \text{ where } n_i^*=
    \begin{cases}
     n_i-1, & \text{if}\ i=j, \\
     n_i, & \text{otherwise}.
      \end{cases}
      \label{Fpay}
\end{equation}
A learner of strategy $i$ adopts the strategy $j$ of the model with a probability given by the Fermi function, as follows

\begin{align}
p_{i,j}=(1+e^{-s(\widehat \pi_j-\widehat \pi_i)})^{-1}.
\label{Fimi}
\end{align}
The parameter $s$ defines the ‘imitation strength’ or 'intensity of selection', i.e. how much $p$ depends on the payoff difference. For $s = 0$, imitation will be entirely random and as $s$ approaches infinity, the imitation process becomes entirely deterministic. In most of the results presented below, we adopt \added{the common value} $s = 1$ \citep{Hilbe2018:PriA}. We also tested other variations of the imitation strength ($s\in \{0.2,1,5\}$)\citep{Okada2017,Krellner2021}.\deleted[id=.]{ and mutation rate ($\mu\in\{0.002, 0.01, 0.05\}$).}

\section{Results}

\subsection{Main Findings}

\begin{figure*}
\centering
\includegraphics[width=1\linewidth]{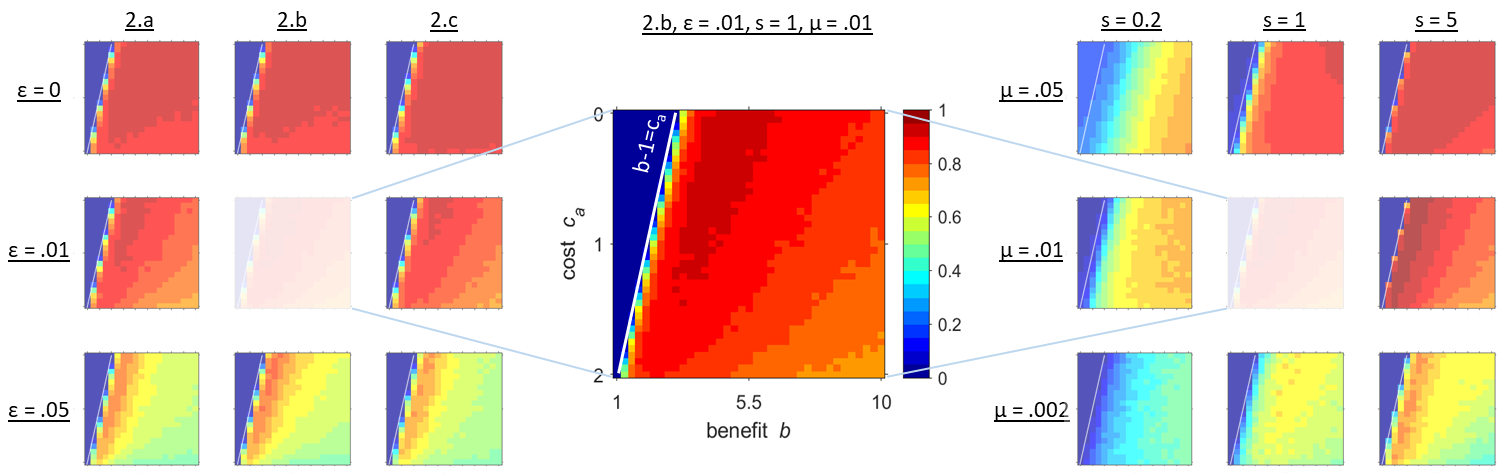}
\caption[Evolution of cooperation]{Evolution of cooperation\textmd{ as a function of the benefit of cooperation ($b$) and cost of commitment arrangement ($c_a$). Stereotypical case (centre) with Assumption 2.b, perception error rate $\epsilon=0.01$, imitation strength $s=1$ and mutation rate $\mu=0.01$. The matrix of graphs on the left shows different game parameters in the same evolutionary setting $s=1$, $\mu=0.01$. The matrix on the right shows different evolutionary parameters in the same game setting $\epsilon=0.01$, Assumption 2.b.} 

Result 'Copop1': \textmd{ The frequency of \textbf{co}operation in the \textbf{pop}ulation is very high for almost the entire area ($>0.7$), where the size of benefits minus the cost of cooperation exceeded the cost of arrangements (i.e. to the right of the diagonal white line $b-1=c_a$).
If errors are frequent, selection weak or mutation  rare, cooperation rates decrease, but cooperation never vanishes.}

Result 'Copop2': \textmd{ Cooperation tends to decrease for higher benefits and smaller costs of arrangements.}}
\label{fig:res1}
\end{figure*}

In our first set of experiments, we simulated the evolutionary process to see if stable cooperation would evolve. We tested a range of values for key parameters, namely the benefit $b$ in the Prisoner's Dilemma (PD) and the cost of arrangement $c_a$. For each data point, we ran 100 simulations as described above and averaged their results. Shown in the centre of Figure \ref{fig:res1} is the result for a standard setting (Assumption 2.b, perception error rate $\epsilon=0.01$, selection strength $s=1$ and mutation rate $\mu=0.01$). High levels of cooperation did evolve almost for the entire parameter space; in particular, wherever the size of the benefits exceeded the combined cost of cooperation and arrangements (see white line $b-1=c_a$). Here, a player who always enters arrangements and always receives benefits can outperform a player who never enters arrangements and never receives benefits.

We tested other settings as well. The left of Figure \ref{fig:res1} shows the results for different game parameters: perception error and assumption. Without perception error (i.e. $\epsilon=0$), cooperation is close to $90\%$ for all assumptions (if $b-1>c_a$). For frequent perception error ($\epsilon=0.05$), cooperation rate is sometimes high but decreases significantly if benefits exceed far beyond $b-1=c_a$. For rare perception error ($\epsilon=0.01$), the results resemble $\epsilon=0$, but the phenomenon of a decrease in cooperation for the highest benefits / lowest arrangement costs is already present. 

Similar results were obtained for different evolutionary parameters. The right of Figure \ref{fig:res1} shows that if the imitation strength and mutation rate are high enough ($s>=1$ and $\mu>=0.01$) cooperation rates are high as well (if $b-1>c_a$). There is also some decrease in cooperation towards high benefits/low arrangement costs. This decrease is reinforced for low mutation rates and high imitation strengths, resembling the effect of high perception error. 

\begin{figure*}
\begin{center}
\includegraphics[width=\linewidth]{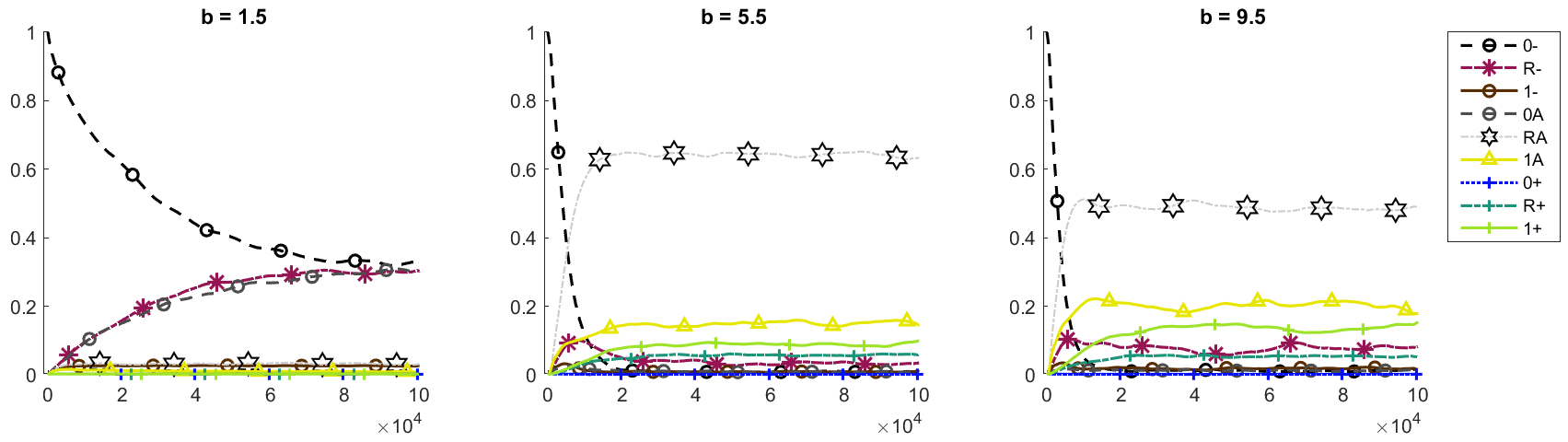}
\caption[Average frequency of all nine strategies over time]{Average frequency of all nine strategies over time. \textmd{Benefit $b$ varies from left to right (low: $b=1.5$, medium: $5.5$, high: $9.5$). In all cases, we set  $c_a=1$, $\epsilon=0.01$; in Assumption 2.b, we also set $s=1$, $\mu=0.01$.} \\
\textmd{For low benefits (left), $b-1>c_a$ does not hold. Cooperation is only profitable if it would not require arrangements. But the only strategy that cooperates and does not enter arrangements, AllC (aka 0+), vanishes as well. The only prevailing strategies are AllD (0- and 0A) and observant faker (R-). What they have in common is  that they do not waste resources on cooperation or arrangements (R- does not trust other fakers or defectors).}

\textmd{For medium benefits (center), RA is by far the most common strategy (}Result 'Strats1'\textmd{), while there are also more naive and more generous cooperators (1A, 1+ and R+) present (}Result 'Strats2'\textmd{).  AllD players are virtually non-existent, but a small share of R- remains (}Result 'Strats3'\textmd{).}

\textmd{For high benefits (right), naive cooperator strategies become even more frequent, and so does R-, lowering the frequency of RA. Note, as Figure \ref{fig:res1} indicates, that increase in fakers' frequency seems to cause overall cooperation to decline, despite the increase in naive cooperators.}
 }
\label{fig3a}
\end{center}
\end{figure*}

\begin{figure*}
\begin{center}
\includegraphics[width=\linewidth]{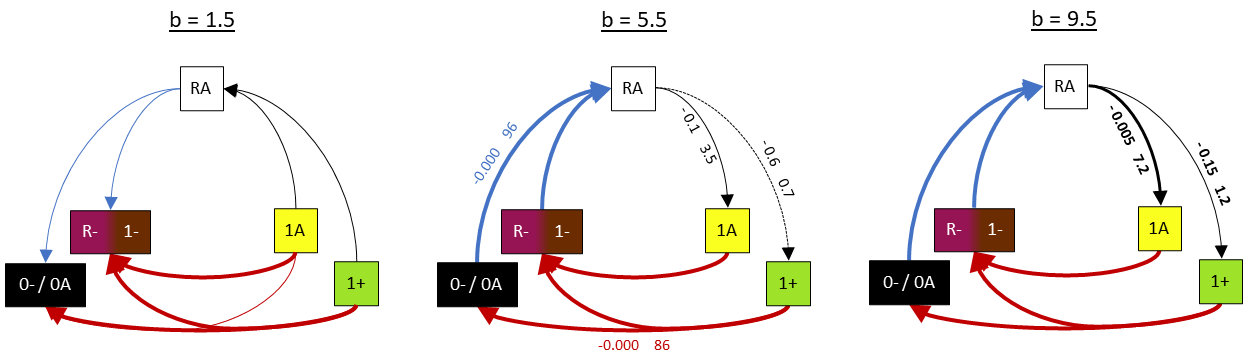}
\caption[Fixation Probabilities under the limit of rare mutations]{Fixation Probabilities under the limit of rare mutations. \textmd{Shown are the most important differences between three benefit cases (as in Figure \ref{fig3a}); namely, from left to right $b=1.5$, $5.5$, $9.5$; Other parameters in all cases: $c_a=1$, $\epsilon=0.01$; In Assumption 2.b, $s=1$, but $\mu\to0$. We ordered strategies in three groups: RA (top), 'nice' (bottom right: 1A, 1+, not shown R+ and 0+) and 'mean' (bottom left: 0-, 0A, R-, 1-, note we combined 0- with 0A, as they behave the same, and R- and 1-, since they are very similar with 1- being occasionally inferior). The direction of the arrows indicates which strategy is risk dominant. 
Negative values show the fixation probability $\rho$ against the direction of the arrow, positive values show it in the direction of the arrow. All values are multitudes of random drift. (For $N=100$, fixation by random drift has a probability of 0.01. A depicted value of 86 indicates a fixation probability of $\rho=0.86$). \\
Dynamics between nice and mean stay mostly the same across benefits. Most mean strategies are strongly favoured to invade nice ones. RA on the other hand is very favoured to invade mean strategies, if $b-1>c_a$ (middle and right). In these conditions, 1A and 1+ are risk dominant against RA (R+ and 0+ are not). While absolute fixation probabilities change little from medium to high benefits, the relative superiority of nice strategies over RA is much more pronounced for high benefits.}}
\label{ch-com:f3b}
\end{center}
\end{figure*}

\begin{figure*}
\begin{center}
\includegraphics[width=\linewidth]{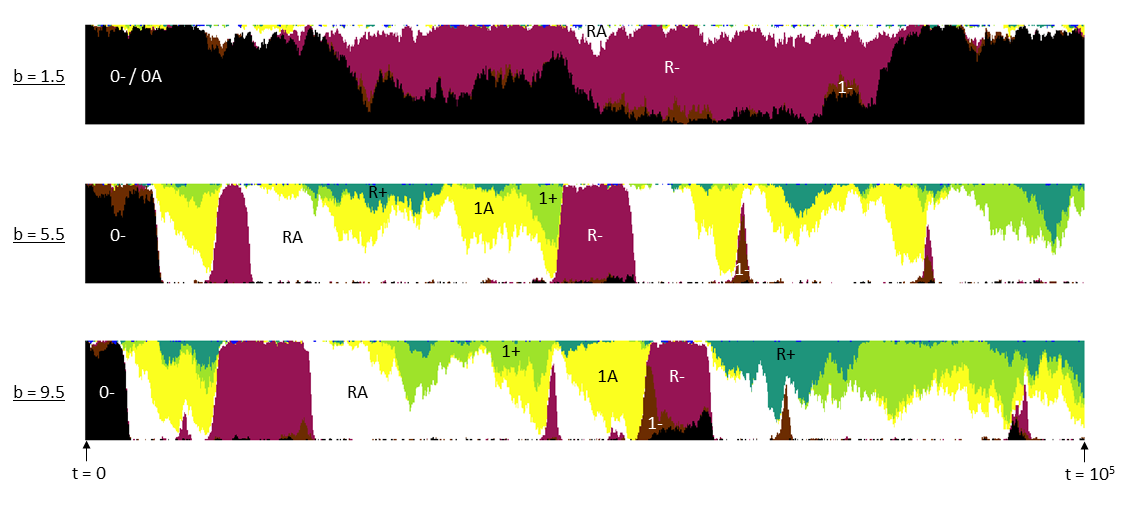}
\caption[Examples of detailed evolutionary dynamics]{Examples of detailed evolutionary dynamics\textmd{, for the three benefit cases (from top to bottom $b=1.5$, $5.5$, $9.5$, setting: $c_a=1$, $\epsilon=0.01$, Assumption 2.b, $s=1$, but $\mu=0.01$). 'Mean' strategies dominate if benefits are low (top), and there seems to be essentially random drift between 0-/0A and R-. For intermediate benefits (center), RA is most common but coexists with changing proportions of nice strategies (1A, 1-, R+). RA becomes less frequent for high benefits (buttom), and nicer strategies make up more of the mixed populations. But invasions of faker strategies (R-, 1-) become more frequent.}
 }
\label{fig3c}
\end{center}
\end{figure*}

Figure \ref{fig3a} shows the evolutionary process over time, averaged at each point in time across $1000$ runs. We showcase three scenarios with different values of the benefit $b$. If benefits are too low ($b=1.5$, left), strategies prevail that never cooperate and never or most rarely commit. Among them is R-, which would commit with good players, but can 'smartly' save the cost of arrangement in this environment. Entering an arrangement only makes sense if one can expect cooperation, which is not the case if there are only other R-, 0- and 0A players (which act like AllD). If benefits are intermediate ($b=5.5$, middle), RA is the most common strategy, but it co-exists with the nice strategies 1A, 1+ and R+, and the faker strategy R-. Both faker strategies (R- and 1-) can profit from the presence of 1+ (as all defectors can), but also from the presence of 1A, since one needs to enter an arrangement with it to trick it into cooperation. R- and 1- do so, but R- is the more frequent of the two. If benefits are too high ($b=9.5$, right), nice strategies become even more frequent, and so do fakers (R- and even 1-). This spread of defecting strategies can explain the decrease in overall cooperation scenarios where benefits are very high compared to the cost of the arrangement $c_a$. 

We can explain some results further with the most important pairwise evolutionary dynamics (Figure \ref{ch-com:f3b}, see SI for a full table of fixation probabilities and description of the standard fixation model). If $b-1>c_a$, a cyclic dynamics pattern is observed between 'mean' strategies (0-, 0A, R-, 1-), RA and 'nice' strategies (1A, 1+, also R+, 0+). RA is risk dominant against all mean strategies, but not against 1A or 1+. \added{Note that, for the so-called risk dominance analysis, we only consider populations in which exactly two strategies are present \citep{Fudenberg2006}. Thus, when considering RA and a nice strategy, there are no fakers to exploit the nice strategy. Consider a population half comprised of RA and half of 1A. Due to perception errors, RA players will sometimes distrust their partner and refuse to enter arrangements with them. This is a mistake, since any player in this population is, in fact, honest and would cooperate if an arrangement was made. Refusing to commit costs RA some benefits that 1A does receive in the same situations (neither type of player can earn benefits if they meet a RA partner who distrusts them).} As $b$ increases, so do RA's relative disadvantages against nice strategies. Both of them (1A and 1+), are in turn always heavily dominated by fakers (and 1+ also by defectors), closing the cycle between RA, nice and mean strategies. 

The narrative can be showcased with three examples of full evolutionary simulations (Figure \ref{fig3c}). The cyclical dynamics can be detected for $b-1>c_a$ , where RA is the first to displace mean strategies, and it is in turn invaded by nice strategies. However, most of the time, RA is not entirely replaced, since the invasion of nice strategies is relatively slow and random. Finally, if RA is entirely replaced, faker strategies are quick to invade and restart the cycle, while sometimes, even residual proportions of RA seem to stop a complete invasion of fakers, cutting the full cycle short. 

\subsection{Validation of Reputation Predictions}

\begin{figure}[H]
\begin{center}
\includegraphics[width=\linewidth]{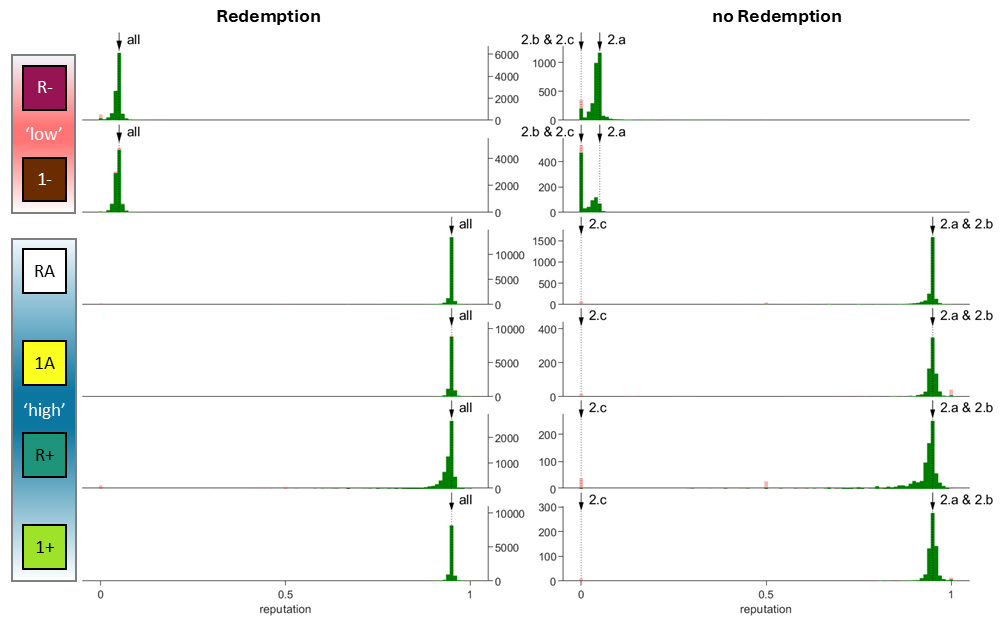}
\caption[Distribution of simulated average reputations]{Distribution of simulated average reputations \textmd{of the six strategies of interest (i.e. for which meaningful average reputations can be predicted). We sampled compositions of $N=100$ players (e.g. 60 RA, 30 1A and 10 R-), for a single perception error $\epsilon=0.05$ (other examples in SI), the same evolutionary parameters ($s=1$, $\mu=0.01$), all three Assumptions 2.a, 2.b and 2.c., three benefits ($b\in\{1.5, 5.5, 9.5\}$) and three costs of arrangement ($c_a\in\{0.25, 1, 1.75\}$), resulting in 27 scenarios in total. For each, we sampled 1000 compositions at random time steps from the evolutionary simulations in that scenarios. Not shown are cases with no R* players in the population, because only those players actually observe interaction and track reputations. If they are not present, reputation does not exist in a meaningful way. This was the case for 1,292 out of 27,000 compositions. All cases are split in right and left depending on whether redemption is possible.} \\ 
\textmd{Redemption is possible (left). All assumptions predict the same, and the predictions were accurate. Virtually all cases where at least three observers are present in the population (green bars) are normally distributed around the predictions. Only for fewer than three observers (light red bars on top of dark green bars), some values differ from predictions, in particular $r=0$ and $r=0.5$.} \\
\textmd{No redemption is possible (right). We had three different assumptions for these cases, 2.a, 2.b or 2.c. Overall, the predictions of 2.a and 2.b seem to fit best. While R- seems to behave more like what 2.a predicted, for 1- Assumption 2.b (and 2.c) makes the better prediction. All other strategies match with the identical predictions of 2.a and 2.b. When there were less than three observers present, we again discover some values which do not fit the prediction of any assumption, at the same conspicuous points $r=0$, $r=0.5$, and also $r=1$. (Some more values at $r=0.66$, $r=0.75$ and $r=0.8$ may have also been artefacts of the number of observers, i.e. three, four and five respectively, but were depicted in green. See R+ with no redemption.)
}}
\label{fig:res2}
\end{center}
\end{figure}

To validate our predictions, we simulated average reputations to compare them with the analytical predictions \citep{Okada2018,Perret123,Fujimoto2023}. For this, we chose a representative sample of population compositions, which we sampled from compositions as they appeared in our evolutionary simulations (i.e. from the simulation Figure \ref{fig:res1} left and middle). We sampled 1000 compositions at random time steps in 81 scenarios (full description in Figure \ref{fig:res2}) and used them for the simulations of reputation described below.

In the simulations, we initiate the image matrix with good opinions, and then run $10^6$ rounds of the \added{game described in subsection \ref{ch:2.1}, consisting of two random players deciding whether to commit and to cooperate in the Prisoner's Dilemma, then assessment by all players.}
We only consider the last half of the rounds to compute the average reputations to minimise the effect of the initial opinion state. Since only R* players care about reputations, only they are considered as observers, and reputations are computed by the average of the opinions of R* players\footnote{Note that all other strategies, namely 0* and 1*, act independently of any opinions they might hold. We therefore chose to exclude their opinions from the simulations to speed them up. This had two consequences. First, in some populations, reputations consist of only a few or even just a single opinion, because there are only a few or even just a single R* player present. This caused some artefacts in cases with less than three observers, as shown in Figure \ref{fig:res2}. Second, we include only meaningful reputations. All the reputations in Figure \ref{fig:res2}\added{ could have potentially impacted payoffs and therefore the evolution of the strategies. Focussing on the opinions of R* strategies seemed, therefore, the most prudent way to test the predictions about reputations.}}\deleted[id=.]{(other entries of the image matrix always remain untouched)}. Note, if there is only one observer, the reputation of a player can only reach the state 0 or 1 at any given time, but can still have an average value between 0 and 1 over time. Reputations of multiple players of the same strategies were averaged as well to obtain final values of mean reputations. 

We plotted the mean reputations for all 1* and R* strategies (since the 0* strategies never commit, their reputation must always stay at 1). Since different perception errors lead to different predictions (prediction 'low': $r=\epsilon$, prediction 'high': $r=1-\epsilon$), we split the values and in Figure \ref{fig:res2} show only $\epsilon=0.05$ and the other results in SI. This high perception error is the most interesting case, since the predictions differ the most between the three assumptions.

In accordance with our assumptions, we further divide the values into two categories. The first includes all compositions where at least one 1* player is present (excluding the focal player itself). In these compositions, we always assume that a player falling to $r=0$ can redeem themselves, since there is at least one naively committing player that would enter into an arrangement with them. In this category, the predictions for all assumptions are the same. Only in the second category, where there are no such naive players to redeem oneself, the predictions between the assumptions differ (2.a keeps the same predictions, 2.b predicts that R- and 1- fall to $r=0$, whereas 2.c predicts that all strategies do so).

Figure \ref{fig:res2} shows the histograms of the simulated average reputations and compares them to the predicted ones (for $\epsilon=0.05$, for other perception errors, see SI). If redemption is possible, almost all values are normally distributed around the universal predictions. If it is not possible, values of R- players match predictions of 2.a most closely, whereas 1- seems better predicted by 2.b and 2.c. For the other strategies, the common prediction of 2.a and 2.b fits best. There are some values which do not fit any assumption, but virtually all of them are cases with two or fewer observers (R* players). These include cases where the entire average reputation consists only of the opinion of one player about one other player. In general, our predictions closely corroborate the simulation results, with 2.a and 2.b providing the best fit. 

\section{Discussion}

\subsection{Summary of Results}

Our results show that reputation and joint commitments (aka arrangements) can sustain cooperation in the Prisoner's Dilemma (which we referred to as Result 'Copop1' in Figure \ref{fig:res1}).
Our evolutionary game theory approach used Monte Carlo simulations that included nine alternative strategies (see Figure \ref{fig:strats} for full list). It showed that cooperation is always frequent, if the benefit of cooperation exceeds the combined cost of cooperation and of making an arrangement. We used an analytical approach inspired by \cite{Fujimoto2022} to predict the reputation to uphold arrangements, and validated these predictions with separate simulations of opinion dynamics \citep{Uchida}. Our results also show that the frequency of cooperation is slightly reduced if benefit sizes become excessive, under the condition that the cost of arrangements is not increased as well (Result 'Copop2'). 

The strategy that only enters an arrangement when its partner is considered good, and that  cooperates only in arrangements, aka \textbf{R}eputation observant \textbf{A}rrangement upholder, or 'RA', is the most successful strategy in our investigation (Result 'Strats1' in Figure \ref{fig3a}). However, it is not risk dominant against some nicer and/or naive strategies, in particular those who are always willing to commit, and either always cooperate \added{(1+)} or at least always cooperate in arrangements \added{(1A)}. Therefore, RA cannot maintain a stable, homogeneous state. Instead, it coexists with significant proportions of these other cooperative strategies (Result 'Strats2'). They in turn invite a yet smaller proportion of faker strategies, which enter arrangements but do not uphold them. The most successful fakers are those who also care about reputation and only enter arrangements with trustworthy players (R-, Result 'Strats3').

\subsection{Limitations}

This investigation is the first to explore the interplay between joint commitment and reputation within an evolutionary game theory framework. As such, its scope was limited and some questions remain. Three topics seem especially interesting: (1) the stability of cooperation under less ideal circumstances, (2) the proof that commitment must be done jointly, and (3) if there are successful assessment rules other than the one studied here.

(1) In our investigation, we simplified the research problem of predicting reputations in two ways. On the one hand, we assumed that all players can observe the interaction. This was also the case in \cite{Fujimoto2022}, on which our approach for reputation predictions is based. On the other hand, we assumed that observers may misperceive the cooperative act but not the presence of an arrangement. If we allow players to sometimes not observe and to make errors about the presence of an arrangement, reputations of cooperative players may suffer, reputations of defectors may increase, and cooperation overall may decrease, comparable to IR under private assessment \citep{Hilbe2018:PriA}. The present investigation did not answer the question if cooperation remains stable if observation is less ideal. 

(2) We conditioned that observers only judge actions in joint commitments, and not if only one partner signalled their willingness to commit. We reasoned that if you were judged every time you commit and your partner does not, your partner had no reason to commit. They could just avoid judgement forever by never committing. Hence, you had no reason to commit either. \deleted[id=.]{Otherwise, if observers judged you, if only your partner did commit, you would have no way to avoid judgement when meeting a person, you do not trust.}\added{It is also problematic if you would be judged if your partner commits but you do not. This way, you could not avoid judgement if you meet a person you do not trust.} Such partners could pressure you with their commitment, even if everybody suspects it to be fake. \added{For these reasons we considered joint commitment to be superior or even essential, and }did not compare our approach with one that judges in such unilateral commitments (see Figure \ref{fig:principles} 'public commitment'). This could be subject to future investigation.

(3) For reputation-based strategies of indirect reciprocity (IR), there exists a multitude of norms \citep{Ohtsuki2006,Okada2020:TwoWays,Perret123}. We can also image other assessment rules for joint commitments. For example, rules that judge honesty in a strict sense, i.e. to dislike it if you did not commit but cooperate anyway. Another example is the opposite of that, i.e. to always appreciate cooperation, even if you did not commit. We did not consider different norms in this investigation, only the one in Figure \ref{fig:principles}). It could be interesting to study different norms, in order to overcome a problem of the current norm; for the strategy presented in this paper, the reputation of a player could become zero, i.e. nobody trusts them any more. Hence, such players might never get the opportunity to join an arrangement again and therefore can not redeem themselves (if there are no naive players). The mentioned appreciation of cooperation outside of arrangements would enable such individuals to redeem themselves. 

The norm introduced in this paper resembles the leading-eight strategy number 7 (L7) \citep{Ohtsuki2006} aka Staying \citep{Sasaki2017,Okada2017,Okada2018}, a very successful indirect reciprocity strategy. The alternative version that appreciates all cooperation resembles L1 aka Standing \citep{Leimar2001}, which is another very successful IR strategy. It seems plausible that its equivalent for joint commitment could be successful as well. However, for all strategies that judge outside of arrangements, studies need to find a way to predict reputations in these more complex settings. The same is true for misperception of arrangements and partial observation. There has been a recent development in analytical models of IR under private assessments which could be suitable for this task \citep{Krellner2023:agr}.

\subsection{Implications and Relevance}

\deleted[id=.]{
The results of this investigation show how joint commitments provided a pathway to the emergence of stable cooperation in non-repeated interactions (Result 'Copop1') and therefore explains how they might have become a part of human behaviour long before key features of civilisation, such as law enforcement and writing, arose. Joint commitment has similar and overlapping concepts. We use 'arrangements' synonymously throughout this paper. Others speak of agreements and mean slightly different things \citep{Gilbert2014,AnhHan,Han}, usually assuming some additional complexity. Even our modern understanding of contracts is a related concept. They can be seen as a legally binding, written joint commitment. Contracts can enable cooperation through additional mechanisms, such as law enforcement. However, our results show that even the most basic form of a joint commitment, which involves only the conditional promises by both parties, has benefits for the parties involved; even without a state (or other powerful third party) to enforce them. It only requires that potential future partners observe who enters joint commitments and who upholds them, or learn about it through gossip \citep{Dunbar1996}. Such conditions were likely in place long before the emergence of modern civilisations.}

\added{
We studied a strategy that uses joint commitment and reputations (Figure \ref{fig:principles}). We show, using evolutionary game theory, how this strategy evolved and established cooperation (main Results 'Copop1' and 'Strats1'). 
We assume that a key element to achieving this is due to the nature of joint commitments. We implemented these commitments as conditional promises (promising only when the other party also promises) that were observable by others. }

\added{
This finding  contributes to contemporary research in three important ways. First, in order to solve the Prisoner's Dilemma, joint commitment does not rely on the enforcement of commitments; it therefore does not require powerful institutions that do the enforcing. This adds to the previously established roles of joint commitment with such enforcement \citep{Han2012:com,Han2013:agreements,Han2022}. Second, this opens up the possibility that joint commitments could have played a larger role during the evolution of the human species. So far, researchers had argued that the role of joint commitment was limited to solving coordination problems \citep{Tomasello2012, Tomasello2019} (see SI for further discussion) presuming that such problems played the most important role in our past, which had been criticised \citep{Forber2015}. Third, our system is an alternative to previous reputation systems for indirect reciprocity that worked without commitment. It provides an important improvement, as it solves the problem of disagreement \citep{Uchida,Panchanathan2003:importance-standing,Krellner2022,Hilbe2018:PriA}. In the next tow paragraphs, we discuss this third implication in more detail.}

\added{First, we need to address the difference between the Prisoner's Dilemma and the so-called Donation Game.} Our results were obtained for the Prisoner's Dilemma, in which both players simultaneously decide to cooperate or defect. \deleted[id=.]{In other versions of this game}\added{In the donation game}, players instead take turns. Player 1 first decides whether to pay the cost (cooperate), which would give player 2 an immediate benefit. \deleted[id=.]{(this is also known as  the donation game). And }\added{During this round, player 2 is just the passive recipient. They can only act in some subsequent round when they are the donor. }In some versions of this game, it is uncertain whether player 2 actually gets the chance to reciprocate, because the players may not meet again. When such a chance to meet again is too small, players must apply indirect reciprocity to maintain cooperation \citep{Schmid2021:DRnIR}. This is why the donation game is most often used to study indirect reciprocity. However, the strategies that work for the donation game also work for the Prisoner's Dilemma.

In contrast, the joint commitment strategy of the current study relied on reciprocity from the current partner. Its principle can therefore not be generalised to the donation game. That is, unless we consider a simple alteration: at the time of joint commitments, the roles of the donor and the recipient are unknown to the players. \deleted[id=.]{We can generalise our findings to such games, since they have}\added{Such a game has the} the same expected value as the standard Prisoner's Dilemma. We therefore argue that the results of our investigation also apply to promises of support. For example, you and I enter an arrangement to support each other in a time of need, without knowing for whom of us the time of need will come. Anecdotally, close friends or spouses implicitly or explicitly enter similar arrangements. Another important example of such arrangements could be pacts between nation states, such as defensive alliances. As for the basic form of joint commitments we envisioned, there is usually no higher power to enforce such arrangements between states. Our system, which combines reputation with joint commitment, is therefore an alternative to previous reputation systems, not in the donation game, but in the Prisoner's Dilemma and its equivalent -- the promise of help in a potential time of need. 

\deleted[id=.]{Finally, when comparing our system with other reputation-systems of indirect reciprocity, two advantages become apparent.}\added{Second, we elaborate why our system can be a better alternative.} The principle of joint commitments gives players the option to defect without penalty, since they are only judged in arrangements, and they themselves can decide whether to enter one. This is a clear advantage over Image Scoring \citep{Panchanathan2003:importance-standing}, which always judges defection \citep{Nowak1998:IRinNature}, and which can only maintain high reputations when it has a high probability of cooperating even with bad players \citep{Schmid2021:DRnIR}. Additionally, it has an advantage over more complex reputation norms (so-called second-order norms). Such norms care about the reputation of the recipient, to avoid bad judgments for defection against bad players \citep{Ohtsuki2006,Okada2020:review}. Such complex reputation norms struggle because any observer may disagree with your opinion that your partner is bad, i.e. under private assessment \citep{Uchida,Hilbe2018:PriA}. IR can still work when utilizing additional mechanisms, such as acting as the observers expect you to act \citep{Krellner2021}\deleted[id=.]{, that requires you to probe and process their opinions, or}, empathy \citep{Radzvilavicius2019}\deleted[id=.]{, in which others try to guess your motives to defect}\added{, gossip \citep{Kessinger2023} or adhering to the opinions of institutions \citep{Radzvilavicius2021:adher}. All of these approaches implement countermeasures to reduce the problem of disagreements. Our approach instead entirely eliminates the need for agreement by removing the observer's opinion as a basis for judging actions, replacing it with the presence of joint commitment.}

\added{We conclude this section with a discussion of the minor results ('Copop2', 'Strats2' and' Strats3'). }While cooperation was frequent most of the time, cooperation rates declined relatively to their peak, when the costs of the arrangements were dwarfed by the benefit of cooperation (Result 'Copop2'). This suggests that it could be beneficial to voluntarily spend extra resources on arrangements, such as a very extravagant wedding, that serves no other purpose than the spending itself. Since the players in our investigation could not choose the amount they (or their partner) would spend on arrangements, we cannot tell whether higher spending can be stable or would be invaded by lower spending. However, it is an interesting direction for future investigations.

In addition to the most common strategy, which applies all the principles of joint commitment and reputation (Result 'Strats1'), there were more naive cooperators present (Result 'Strats2'). Mapping these findings to the real world, we may expect a similar polymorphism in humans. Most people are probably careful with whom they enter arrangements and will uphold their own commitments. But some may enter arrangements more often than they have reasons to do so (i.e. they are more trusting), like the second most common strategy in our investigation (1A). Given that most people are trustworthy and will cooperate in arrangements, their naive commitments may, on average, pay off.

We may also expect a significant proportion of people to enter arrangements with the intention of breaking them (Result 'Strats3'). The best way to do that might be to also carefully observe the interaction of others and only try to trick those who are worth tricking, since those will cooperate after arrangements. On a very cynical note, such fakers could deploy an additional strategy and smear others' reputations. Because if the victims of such gossip find fewer and fewer people willing to enter arrangements with them, they rely more on arrangements with the fakers. This is another interesting direction for future research.

\section*{Acknowledgements} 
During the preparation of this work, the authors used Writefull and LanguageTool in order to ensure correct English and improve readability. After using these tools, the authors reviewed and edited the content as needed and take full responsibility for the content of the publication.

T.A.H. is supported by EPSRC (grant EP/Y00857X/1).  

\bibliographystyle{apalike}
\bibliography{main}


\end{document}


\maketitle


\section{Additional Results}


\begin{table}[!ht]
    \centering
    \caption[All fixation probabilities for results from Figure \ref{ch-com:f3b}, $b=1.5$]{Fixation probabilities for results from the main text, $b=1.5$}
    \begin{tabular}{|l|l|l|l|l|l|l|l|l|l|}
    \hline 
        ~ & ~ & ~ & ~ & ~ & ~ & resident & ~ & ~ & ~ \\ \hline
        ~ & ~ & 1- & R- & 0- & 1A & RA & 0A & 1+ & R+ \\ \hline
        ~ & 1- & ~ & 0.00\% & 0.00\% & 63.77\% & 0.00\% & 0.00\% & 63.77\% & 12.10\% \\ \hline
        ~ & R- & 7.85\% & ~ & 1.00\% & 86.24\% & 5.31\% & 1.00\% & 86.38\% & 63.96\% \\ \hline
        ~ & 0- & 7.42\% & 1.00\% & ~ & 5.36\% & 5.31\% & 1.00\% & 63.97\% & 63.96\% \\ \hline
        ~ & 1A & 0.00\% & 0.00\% & 0.00\% & ~ & 0.78\% & 0.00\% & 1.00\% & 1.19\% \\ \hline
        invader & RA & 0.00\% & 0.00\% & 0.00\% & 1.26\% & ~ & 0.00\% & 2.02\% & 2.31\% \\ \hline
        ~ & 0A & 7.42\% & 1.00\% & 1.00\% & 5.36\% & 5.31\% & ~ & 63.97\% & 63.96\% \\ \hline
        ~ & 1+ & 0.00\% & 0.00\% & 0.00\% & 1.00\% & 0.36\% & 0.00\% & ~ & 0.59\% \\ \hline
        ~ & R+ & 0.00\% & 0.00\% & 0.00\% & 0.90\% & 0.31\% & 0.00\% & 1.56\% & ~ \\ \hline
    \end{tabular}
\end{table}

\begin{table}[!ht]
    \centering
    \caption[All fixation probabilities for results from Figure \ref{ch-com:f3b}, $b=5.5$]{Fixation probabilities for results from the main text, $b=5.5$}
    \begin{tabular}{|l|l|l|l|l|l|l|l|l|l|}
    \hline
        ~ & ~ & ~ & ~ & ~ & ~ & resident & ~ & ~ & ~ \\ \hline
        ~ & ~ & 1- & R- & 0- & 1A & RA & 0A & 1+ & R+ \\ \hline
        ~ & 1- & ~ & 0.00\% & 0.00\% & 65.20\% & 0.00\% & 0.00\% & 65.20\% & 14.56\% \\ \hline
        ~ & R- & 7.85\% & ~ & 1.00\% & 86.78\% & 0.00\% & 1.00\% & 86.92\% & 65.38\% \\ \hline
        ~ & 0- & 7.42\% & 1.00\% & ~ & 0.00\% & 0.00\% & 1.00\% & 65.38\% & 65.38\% \\ \hline
        ~ & 1A & 0.00\% & 0.00\% & 96.87\% & ~ & 3.51\% & 96.87\% & 1.00\% & 3.48\% \\ \hline
        i. & RA & 96.64\% & 96.64\% & 96.64\% & 0.12\% & ~ & 96.64\% & 0.63\% & 2.37\% \\ \hline
        ~ & 0A & 7.42\% & 1.00\% & 1.00\% & 0.00\% & 0.00\% & ~ & 65.38\% & 65.38\% \\ \hline
        ~ & 1+ & 0.00\% & 0.00\% & 0.00\% & 1.00\% & 0.74\% & 0.00\% & ~ & 0.59\% \\ \hline
        ~ & R+ & 0.00\% & 0.00\% & 0.00\% & 0.36\% & 0.30\% & 0.00\% & 1.56\% & ~ \\ \hline
    \end{tabular}
\end{table}

\begin{table}[!ht]
    \centering
    \caption[All fixation probabilities for results from Figure \ref{ch-com:f3b}t, $b=9.5$]{Fixation probabilities for results from the main text, $b=9.5$}
    \begin{tabular}{|l|l|l|l|l|l|l|l|l|l|}    
    \hline
        ~ & ~ & ~ & ~ & ~ & ~ & resident & ~ & ~ & ~ \\ \hline
        ~ & ~ & 1- & R- & 0- & 1A & RA & 0A & 1+ & R+ \\ \hline
        ~ & 1- & ~ & 0.00\% & 0.00\% & 66.58\% & 0.00\% & 0.00\% & 66.58\% & 17.11\% \\ \hline
        ~ & R- & 7.85\% & ~ & 1.00\% & 87.30\% & 0.00\% & 1.00\% & 87.44\% & 66.74\% \\ \hline
        ~ & 0- & 7.42\% & 1.00\% & ~ & 0.00\% & 0.00\% & 1.00\% & 66.74\% & 66.74\% \\ \hline
        ~ & 1A & 0.00\% & 0.00\% & 99.94\% & ~ & 7.15\% & 99.94\% & 1.00\% & 6.94\% \\ \hline
        i. & RA & 99.93\% & 99.93\% & 99.93\% & 0.00\% & ~ & 99.93\% & 0.15\% & 2.43\% \\ \hline
        ~ & 0A & 7.42\% & 1.00\% & 1.00\% & 0.00\% & 0.00\% & ~ & 66.74\% & 66.74\% \\ \hline
        ~ & 1+ & 0.00\% & 0.00\% & 0.00\% & 1.00\% & 1.18\% & 0.00\% & ~ & 0.59\% \\ \hline
        ~ & R+ & 0.00\% & 0.00\% & 0.00\% & 0.10\% & 0.28\% & 0.00\% & 1.56\% & ~ \\ \hline
    \end{tabular}
\end{table}

\newpage

\begin{figure*}
\begin{center}
\includegraphics[width=\linewidth]{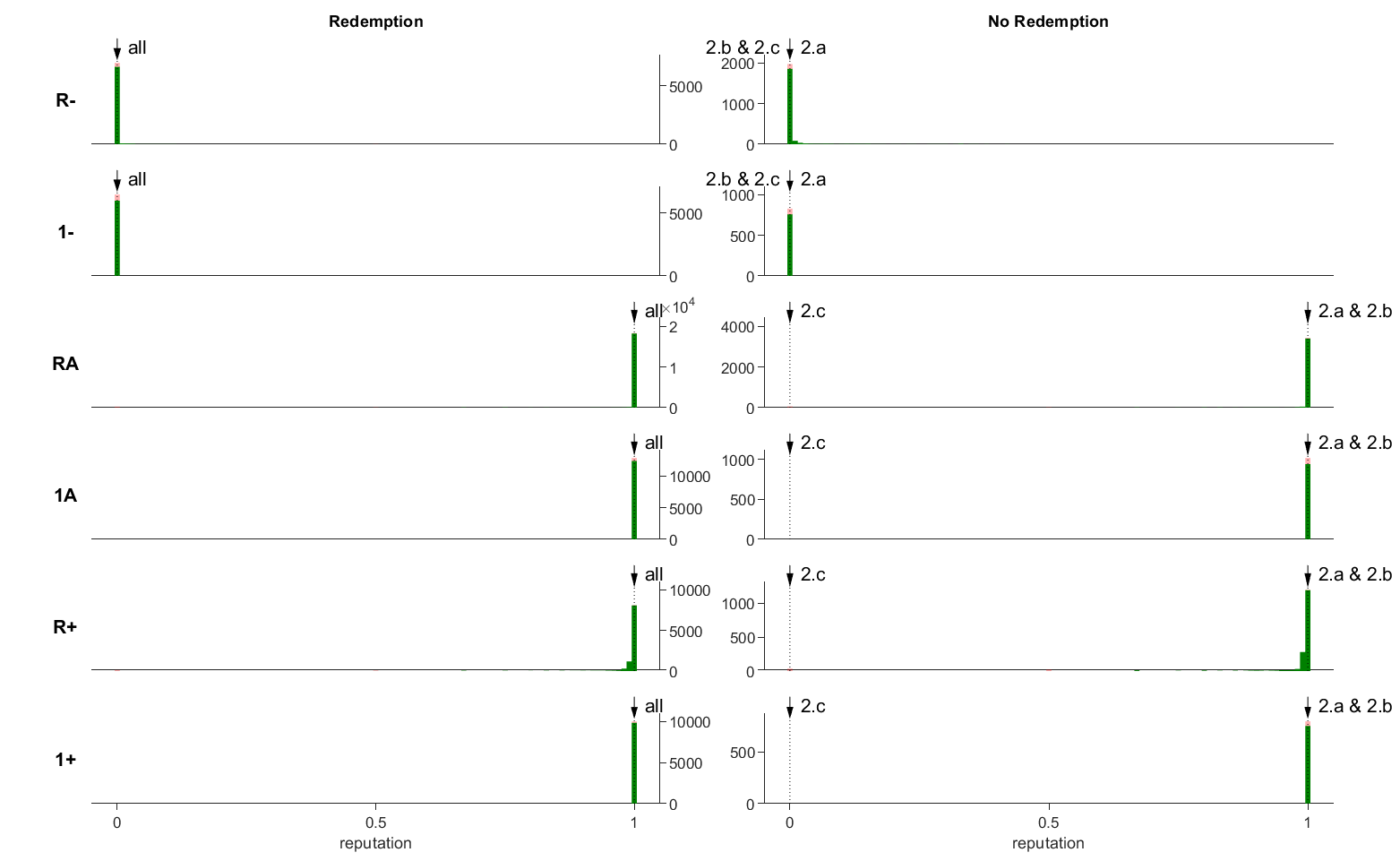}
\caption[Distribution of simulated average reputations, $\epsilon=0$]{Distribution of simulated average reputations,  $\epsilon=0$, \textmd{of the six strategies of interest (i.e. for which meaningful average reputations can be predicted). We sampled compositions of $N=100$ players (e.g. 60 RA, 30 1A and 10 R-), for a single perception error $\epsilon=0.05$ (other examples in SI), the same evolutionary parameters ($s=1$, $\mu=0.01$), all three assumptions 2.a, 2.b and 2.c., three benefits ($b\in\{1.5, 5.5, 9.5\}$) and three costs of arrangement ($c_a\in\{0.25, 1, 1.75\}$), resulting in 27 scenarios in total. For each, we sampled 1000 compositions at random time steps from the evolutionary simulations in that scenarios. 
}
 }
\label{S-fig1}
\end{center}
\end{figure*}

\begin{figure*}
\begin{center}
\includegraphics[width=\linewidth]{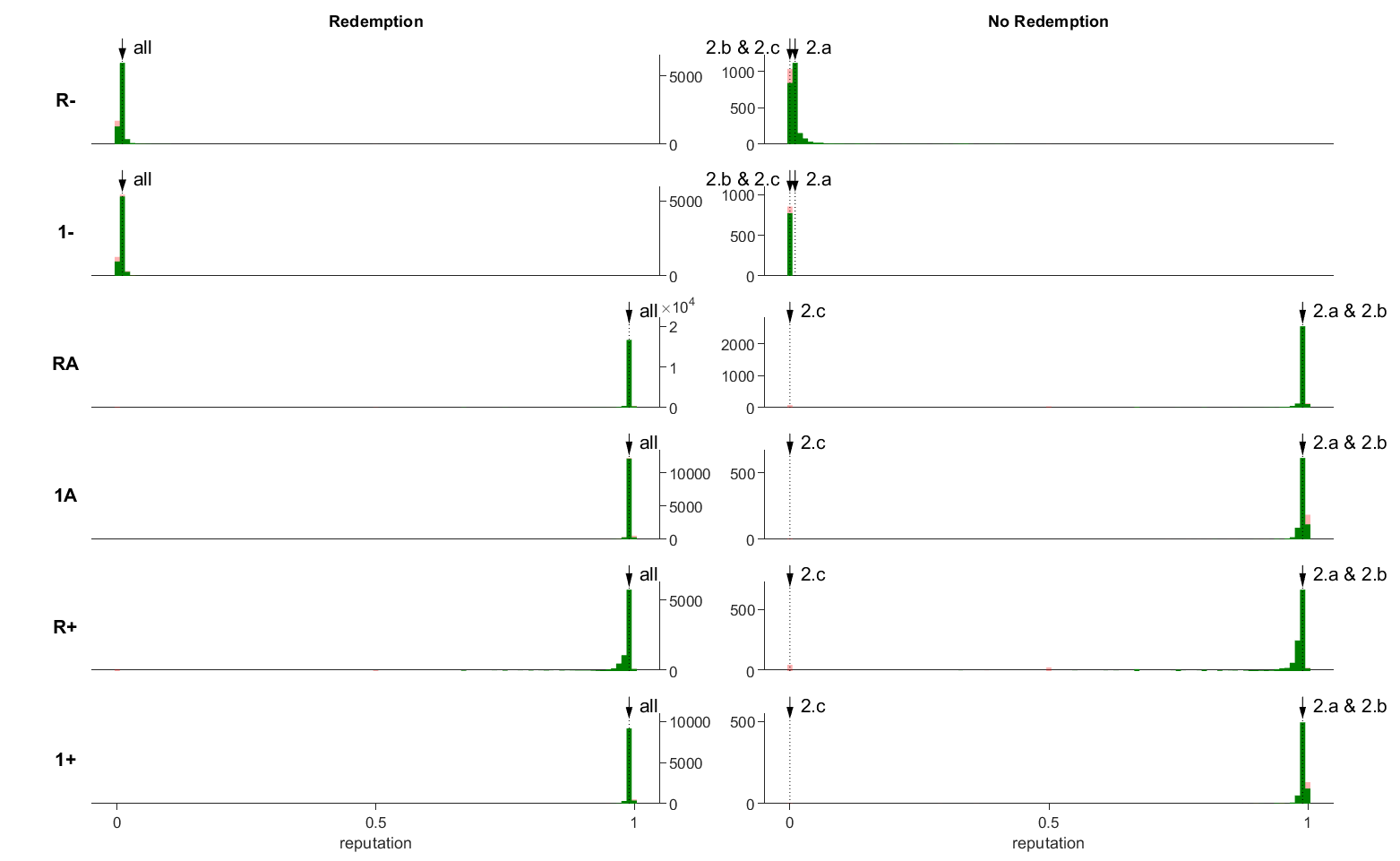}
\caption[Distribution of simulated average reputations, $\epsilon=0.01$]{Distribution of simulated average reputations, $\epsilon=0.01$, \textmd{of the six strategies of interest (i.e. for which meaningful average reputations can be predicted). We sampled compositions of $N=100$ players (e.g. 60 RA, 30 1A and 10 R-), for a single perception error $\epsilon=0.05$ (other examples in SI), the same evolutionary parameters ($s=1$, $\mu=0.01$), all three assumptions 2.a, 2.b and 2.c., three benefits ($b\in\{1.5, 5.5, 9.5\}$) and three costs of arrangement ($c_a\in\{0.25, 1, 1.75\}$), resulting in 27 scenarios in total. For each, we sampled 1000 compositions at random time steps from the evolutionary simulations in that scenarios. 
}
 }
\label{S-fig2}
\end{center}
\end{figure*}

\section{Additional Discussion}

\subsection{Coordination Problems}

In the main text we mentioned that research on joint commitment from fields such as philosophy and psychology had not considered joint commitment to play a role in solving problems such as the Prisoner's Dilemma \citep{Doebeli2005}. Instead, it was argued that joint commitments are used to solve coordination problems \citep{Tomasello2012}. Our investigation did not have the scope to answer the question of the ultimate function of joint commitment indefinitely, and it is very possible that its function is both to solve coordination problems as well as social dilemmas, but we will discuss this distinction a bit further.

\begin{figure}
\centering
  \includegraphics[width=0.8\linewidth]{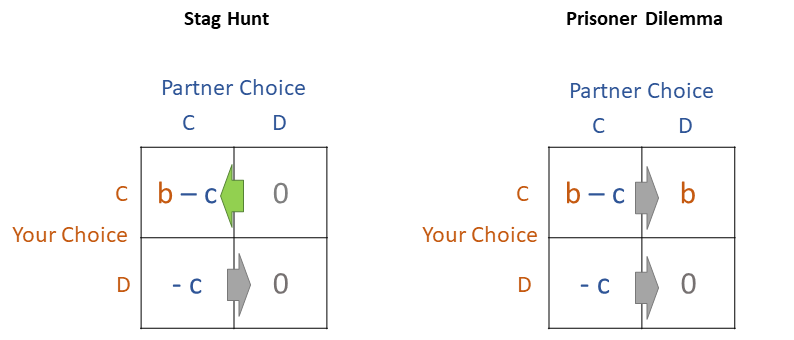}
  \caption[Payoff matrix for your partner in stag hunt game and Prisoner's Dilemma]{Payoff matrix for your partner in stag hunt game and Prisoner's Dilemma\textmd{, version with only two variables: benefit $b$ and cost $c$, but $b>c$ (otherwise mutual defection is better than mutual cooperation in both games). In the stag hunt game, your partner's preferred action depends on your action. If you cooperate (C), they are tempted to cooperate as well. The same is true if you can make them believe that you will cooperate \added{by commitment}. In contrast, in the Prisoner's Dilemma (PD), your partner is always tempted to defect, no matter what they believe you will do.}}
  \label{fig1}
  \end{figure}
 
Both ideas assume that joint commitments are important because they help us achieve common goals. 
Achieving common goals often requires solving coordination problems. A coordination problem occurs, for example, when two people consider hunting a stag. Neither can catch the stag alone,  but if they manage to coordinate and hunt together, they will achieve their goal and receive the reward. If they fail to coordinate, one of them may hunt alone, fail, and waste effort for no reward. This is illustrated in Figure \ref{fig1}. Here, your commitment can solve the coordination problem easily. If you promise to hunt for the stag, your partner's best choice is to join in.

Achieving a common goal gets harder in the presence of temptation \citep{Forber2015,Kachel2019}. Let us consider a stag hunt game with temptation. Instead of hunting a stag today, you could eat some stashed food. It is not enough to feed you and your partner, so this would be less valuable than the expected reward for hunting the stag. But if you can get your partner to go away and hunt, you can have it all for yourself. This is the best outcome for you, and the worst outcome for your partner. This scenario is an instance of 
the Prisoner's Dilemma. 
The payoff in comparison to the original stag hunt case is shown in Figure \ref{fig1}. In an unaltered Prisoner’s Dilemma, the logical outcome is for both to defect, to avoid the worst outcome of cooperating while the partner defects. And crucially, your commitment does not change your partner's preference to defect.

\subsection{Relation to Implicit Commitments}

Humans might have become such experts in joint commitments \citep{Tomasello2019}, that they can even handle implicit commitments. For example, when we sit around a fire, I might say “I am going to gather more wood”. You may say: “I will come with you.”, hence committing explicitly to help me. But you may also just stand up and walk towards me or towards the forest. Your commitment to our joint task is very subtle. A person, who rarely or never gets wood, is frowned upon, which may become apparent after a few nights. But, I would be immediately irritated, if you stood up and came, but then just left halfway there, or if you just trotted along without ever lifting a single twig. This is, because I sensed an implicit commitment, for which I hold you accountable. 

However, such implicit joint commitments may be hard for others to judge. They therefore might work only for direct reciprocity, in which one interacts with the same partner repeatedly and judges them on one's own interactions with them instead of (also) the interactions they have with others. Such a system exists in our model, if there are only 2 players that enter arrangements. But in this case, cooperation might cease forever, if one of them loses trust. 

To enable a reputation-based system of indirect reciprocity, certain rituals of explicit joint commitment might have emerged. These rituals might not be needed as signals to one's partner, who would likely understand the commitment implicitly. An example is marriage, which has many rich traditions that seem flamboyant and wasteful. Many couples may ask, what benefit these expenses and efforts have for them, for their commitment and their love. But our point of view on the matter would be, that the target of these rituals are the people around them. The goal is to make the joint commitment as clear and public as possible. But with the ultimate goal of maximizing your partner's risk for breaking this commitment in the future.


\subsection{Why we focus on trusting strategies}
\label{SI:trust}

\added{For the game or for real life, we can imagine either initial state of opinion towards an unknown individual: to trust (because I have no reason to distrust) or to distrust (because I have no reason to trust either). In our study, we decided that players would always start with trust. There are two reasons for that. First, because it reduces the scope and complexity of the current study. Second, because of the consequences of trust and distrust, as discussed below.}

\added{Starting in a state of trust is ultimately a part of the strategy. We can imagine many different types of RA strategies. For example, one that starts with good opinions about their co-players, one that starts with random opinions and one that starts bad opinions. We simply chose to only study the strategies that start with good opinions, leaving the other versions to be done by future research.} 

\added{What are the likely consequences of trust vs distrust? Keep in mind, that we are interested in the long-term behaviour. Would the initial state influence the long-term behaviour? That depends. If there is at least some trust, players will start entering joint commitments. After that, trust is updated. Recall that we analytically predicted and verified with simulations, that reputations have only a few expected values ('low' and 'high', as well as 'zero' and 'none'). If there is at least some trust, which includes opinions to be initially random, these predictions do not change.}

\added{If initially there is only distrust, however, things are different. Remember Assumption 2, that player need to be able to enter arrangements at least some of the time. If all R*-players are initially distrusting, and no 1*-players are present, there cannot be any arrangements. We therefore face the results of Assumption 2.c, where all  players have a reputation of $r=0$ and will keep that reputation forever. But instead of being in that state after a very long period of time in the case of initial trust, they would start in that state in the case of initial distrust. Starting in a state of total distrust therefore has sudden catastrophic consequences for cooperation of players that only cooperate in arrangements (*A-player).} 

\added{However, there is a silver lining. Let us reiterate: this is only the case, if there are no 1*-players. And let us point out further that we already included these cases in our model. Initial distrust is not different from Assumption 2.c, which we already studied. We already considered that cooperation of *A-players ceases in those populations, since we only concerned the long-term behaviour. Starting at this point as well as ending up in it does not change the results.} 

\added{So, even though we did not explicitly study versions of our strategy that start with different states than total trust, we have the strong suspicion, that our results would hold for any other initial states, or at least for those that do not cause all reputations to immediately and unredeemedly collapse.}

\bibliographystyle{apalike}
\bibliography{main}


